\documentclass{80SA}
\usepackage{times}

\title{%
Multipole correlations of $t_{\rm 2g}$-orbital Hubbard model
with spin-orbit coupling}

\author{%
Hiroaki Onishi\thanks{E-mail address: onishi.hiroaki@jaea.go.jp}
}
\inst{%
Advanced Science Research Center,
Japan Atomic Energy Agency,
Tokai, Ibaraki 319-1195, JAPAN
}
\abst{%
We investigate the ground-state properties of
a one-dimensional $t_{\rm 2g}$-orbital Hubbard model
including an atomic spin-orbit coupling
by using numerical methods,
such as Lanczos diagonalization and density-matrix renormalization group.
As the spin-orbit coupling increases,
we find a ground-state transition
from a paramegnetic state to a ferromagnetic state.
In the ferromagnetic state,
since the spin-orbit coupling mixes spin and orbital states
with complex number coefficients,
an antiferro-orbital state with complex orbitals appears.
According to the appearance of the complex orbital state,
we observe an enhancement of $\Gamma_{4u}$ octupole correlations.
}

\kword{%
$t_{\rm 2g}$ orbitals,
spin-orbit coupling,
multipole,
density-matrix renormalization group
}

\begin{document}
\maketitle


The competition and cooperation between spin and orbital degrees of freedom
in strongly correlated electron systems
manifest itself in the emergence of various types of
spin-orbital ordered and quantum liquid phases.
\cite{orbital2001,Tokura2000,Hotta2006}
In general, among competing interactions involving spin and orbital,
the spin-orbit coupling is supposed to be weak
in $3d$ transition-metal oxides such as cupurates and manganites,
while as we move to $4d$ and $5d$ electrons,
the spin-orbit coupling becomes strong
and responsible for magnetic, transport, and optical properties.
When the spin-orbit coupling is dominant,
spin and orbital are not independent,
but instead the total angular momentum gives
a good description of the many-body state.
In fact,
it has been suggested that Sr$_{2}$IrO$_{4}$,
in which Ir$^{4+}$ ions have five electrons
in triply degenerate $t_{\rm 2g}$ orbitals,
exhibits a novel Mott-insulating state
with an effective total angular momentum $J_{\rm eff}$=$1/2$
due to a strong spin-orbit coupling.
\cite{Kim2008,Kim2009, Chikara2009,Moon2009}
In the limit of strong spin-orbit coupling,
the ground-state Kramers doublet at a local ion
can be described by an isospin with $J_{\rm eff}$=$1/2$.
\cite{Jin2009,Jackeli2009}
The exchange interaction among isospins can lead to
a variety of ordering and fluctuation phenomena
of spin-orbital entangled states.

When we move to heavy-element $f$-electron systems
such as rare-earth and actinide compounds,
the spin-orbit coupling is large
comparing with other energy scales.
In such a case, we usually classify the complicated spin-orbital state
from the viewpoint of multipole,
which is described by the total angular momentum.
Indeed, the multipole physics has been actively discussed
in the field of heavy electrons.
\cite{Kuramoto2009}
A recent trend is to unveil exotic high-order multipole ordering.
As an attempt to clarify multipole properties of $f$-electron systems
from a microscopic viewpoint,
we have numerically studied multipole correlations
of an $f$-orbital Hubbard model
on the basis of the $j$-$j$ coupling scheme.
\cite{Onishi2006}
We believe that it is also important to clarify multipole properties
in $d$-electron systems under the effect of the spin-orbit coupling.

In this paper, we investigate multipole properties in the ground state
of a one-dimensional $t_{\rm 2g}$-orbital Hubbard model
including the spin-orbit coupling by numerical methods.
With increasing the spin-orbit coupling,
the ground state changes from a paramagnetic state
to a ferromagnetic state
in terms of the magnitude of the total spin.
In the ferromagnetic phase,
antiferro-dipole correlations develop
even when the spin state is ferromagnetic
due to the orbital contribution.
On the other hand, the spin-orbit coupling induces
a complex orbital state,
in which real $xy$, $yz$, and $zx$ orbitals are mixed
with complex number coefficients.
According to the complex orbital state,
$\Gamma_{4u}$ octupole correlations are enhanced.


Let us consider triply degenerate $t_{\rm 2g}$ orbitals
on a one-dimensional chain along the $x$ direction
with five electrons per site.
The one-dimensional $t_{\rm 2g}$-orbital Hubbard model
with the spin-orbit coupling is described by
\begin{eqnarray}
\label{eq:ham}
 H  \! \! \! \! \! &=&  \! \! \! \! \!
 \sum_{i,\tau,\tau',\sigma}
 t_{\tau\tau'}
 (d_{i\tau\sigma}^{\dag} d_{i+1\tau'\sigma}+{\rm h.c.})
 +\lambda\sum_{i} {\bf L}_{i} \cdot {\bf S}_{i}
 \nonumber\\
 &&  \! \! \! \! \!
 +U \sum_{i,\tau}
 \rho_{i\tau\uparrow} \rho_{i\tau\downarrow}
 +(U'/2) \sum_{i,\sigma,\sigma',\tau\neq\tau'} 
 \rho_{i\tau\sigma} \rho_{i\tau'\sigma'}
 \nonumber\\
 && \! \! \! \! \!
 +(J/2) \sum_{i,\sigma,\sigma',\tau\neq\tau'} 
 d_{i\tau\sigma}^{\dag} d_{i\tau'\sigma'}^{\dag}
 d_{i\tau\sigma'} d_{i\tau'\sigma}
 \nonumber\\
 && \! \! \! \! \!
 +(J'/2) \sum_{i,\sigma\neq\sigma',\tau\neq\tau'} 
 d_{i\tau\sigma}^{\dag} d_{i\tau\sigma'}^{\dag}
 d_{i\tau'\sigma'} d_{i\tau'\sigma},
\end{eqnarray}
where $d_{i\tau\sigma}$ ($d_{i\tau\sigma}^{\dag}$)
is an annihilation (creation) operator for an electron
with spin $\sigma$ (=$\uparrow,\downarrow$)
in orbital $\tau$ (=$xy,yz,zx$)
at site $i$,
and $\rho_{i\tau\sigma}$=$d_{i\tau\sigma}^{\dag}d_{i\tau\sigma}$.
The hopping amplitude is given by $t_{xy,xy}$=$t_{zx,zx}$=$t$
and zero for other combinations of orbitals.
Hereafter, $t$ is taken as the energy unit.
${\bf L}_i$ and ${\bf S}_i$ represent
orbital and spin angular momentum operators, respectively,
and $\lambda$ is the spin-orbit coupling.
$U$, $U'$, $J$, and $J'$ denote
intra-orbital Coulomb, inter-orbital Coulomb, exchange,
and pair-hopping interactions, respectively.
We assume
$U$=$U'$+$J$+$J'$
due to the rotation symmetry in the local orbital space
and $J'$=$J$
due to the reality of the orbital function.
\cite{Dagotto2001}
Throughout this paper,
we set $\hbar$=$k_{\rm B}$=$1$.

We investigate the ground-state properties of the model (\ref{eq:ham})
by exploiting a finite-system
density-matrix renormalization group (DMRG) method
with open boundary conditions.
\cite{White1992}
The number of states kept for each block is up to $m$=$120$,
and the truncation error is estimated to be $10^{-4}$$\sim$$10^{-5}$.
We remark that due to the three orbitals in one site,
the number of bases for the single site is 64,
and the size of the superblock Hilbert space grows as
$m^2$$\times$$64^2$.
To reduce the size of the Hilbert space,
we usually decompose the Hilbert space into a block-diagonal form
by using symmetries of the Hamiltonian.
In the present case,
however,
the spin-orbit coupling breaks the spin SU(2) symmetry,
so that we cannot utilize $S_{\rm tot}^z$ as a good quantum number,
where $S_{\rm tot}^z$ is the $z$ component of the total spin.
Since we ignore $e_{\rm g}$ orbitals among $d$ orbitals,
the total angular momentum is not a conserved quantity.
The total number of electrons can be used as a good quantum number.
Thus, since DMRG calculations consume much CPU times,
we supplementally use a Lanczos diagonalization method
for the analysis of a four-site periodic chain
to accumulate results with relatively short CPU times.


\begin{figure}[t]
\begin{center}
\includegraphics[width=0.95\linewidth]{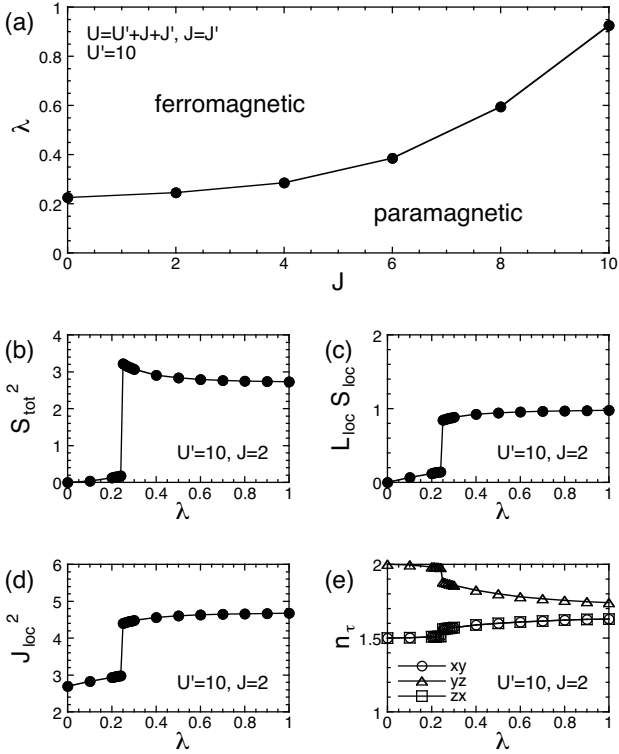}
\end{center}
\caption{%
Lanczos results for the four-site periodic chain.
(a) The ground-state phase diagram
in the ($J$,$\lambda$) plane for $U'$=$10$.
(b) The magnitude of the total spin in the whole system.
(c) The correlation between spin and orbital in the single site.
(d) The magnitude of the total angular momentum in the single site.
(e) The charge density in each orbital.
}
\label{f1}
\end{figure}

Let us first look at Lanczos results for the four-site periodic chain.
In Fig.~1(a),
we show the phase diagram in the ($J$,$\lambda$) plane for $U'$=$10$.
The phase boundary is determined by the magnitude of
the total spin ${\bf S}_{\rm tot}^{2}$.
As shown in Fig.~1(b),
${\bf S}_{\rm tot}^{2}$ is almost zero for small $\lambda$,
indicating a spin-singlet ground state.
As $\lambda$ increases,
we find a transition to a ferromagnetic state
with finite ${\bf S}_{\rm tot}^{2}$.
Note that even in the limit of large $\lambda$,
${\bf S}_{\rm tot}^{2}$ does not approach the maximum value 2(2+1)=6,
since the spin-orbit coupling mixes spin up and down states
and the complete ferromagnetic state is disturbed.
In Fig.~1(c),
we plot the correlation between spin and orbital in the local site
${\bf L}_{\rm loc}\cdot{\bf S}_{\rm loc}$.
At $\lambda$=$0$, there is no correlation between spin and orbital.
As $\lambda$ increases,
the spin-orbital correlation develops and approaches one
in the limit of large $\lambda$,
indicating totally parallel spin and orbital angular momenta.
In Fig.~1(d),
the magnitude of the total angular momentum in the single site
${\bf J}_{\rm loc}^{2}$ is shown.
At the transition point,
${\bf J}_{\rm loc}^{2}$ exhibits a sudden increase,
since the spin-orbit coupling stabilizes
a large total angular momentum state at every local sites.
Note again that ${\bf J}_{\rm loc}^{2}$ does not reach the maximum value
$\frac{5}{2}(\frac{5}{2}+1)$=$\frac{35}{4}$
in the limit of large $\lambda$,
since the total angular momentum is not a conserved quantity.
Regarding the orbital state,
we show the charge density in each orbital in Fig.~1(e).
Due to the spatial anisotropy of orbitals,
one hole is preferably accommodated in itinerant $xy$ or $zx$ orbitals
in each site,
while localized $yz$ orbitals are doubly occupied.
Measuring charge correlations,
we find that holes occupy real $xy$ or $zx$ orbital alternately
for small $\lambda$ (not shown).
Namely, the ground state is a {\it real} orbital state.
For large $\lambda$, however,
$xy$, $yz$, and $zx$ orbitals are mixed with complex number coefficients
by the spin-orbit coupling,
leading to a {\it complex} orbital state.

\begin{table}[t]
\begin{center}
\begin{tabular}{ll}
\hline
$\Gamma_{\gamma}$ multipole & multipole operator \\
\hline
$\Gamma_{4u}$ dipole
& $J_x$, $J_y$, $J_z$
\\
$\Gamma_{3g}$ quadrupole
& $O_{u}$=(1/2)(2$J_z^2$$-$$J_x^2$$-$$J_y^2$)
\\
& $O_{v}$=($\sqrt{3}$/2)($J_x^2$$-$$J_y^2$)
\\
$\Gamma_{5g}$ quadrupole
& $O_{yz}$=($\sqrt{3}$/2)$\overline{J_y J_z}$
\\
& $O_{zx}$=($\sqrt{3}$/2)$\overline{J_z J_x}$
\\
& $O_{xy}$=($\sqrt{3}$/2)$\overline{J_x J_y}$
\\
$\Gamma_{2u}$ octupole
& $T_{xyz}$=($\sqrt{15}$/6)$\overline{J_x J_y J_z}$
\\
$\Gamma_{4u}$ octupole
& $T_{x}^{\alpha}$=(1/2)(2$J_x^3$$-$$\overline{J_x J_y^2}$$-$$\overline{J_z^2 J_x}$)
\\
& $T_{y}^{\alpha}$=(1/2)(2$J_y^3$$-$$\overline{J_y J_z^2}$$-$$\overline{J_x^2 J_y}$)
\\
& $T_{z}^{\alpha}$=(1/2)(2$J_z^3$$-$$\overline{J_z J_x^2}$$-$$\overline{J_y^2 J_z}$)
\\
$\Gamma_{5u}$ octupole
& $T_{x}^{\beta}$=($\sqrt{15}$/6)($\overline{J_x J_y^2}$$-$$\overline{J_z^2 J_x}$)
\\
& $T_{y}^{\beta}$=($\sqrt{15}$/6)($\overline{J_y J_z^2}$$-$$\overline{J_x^2 J_y}$)
\\
& $T_{z}^{\beta}$=($\sqrt{15}$/6)($\overline{J_z J_x^2}$$-$$\overline{J_y^2 J_z}$)
\\
\hline
\end{tabular}
\end{center}
\caption{%
Definition of multipole operators up to rank 3.
The overline on the product denotes the operation of taking
all possible permutations in terms of cartesian components,
e.g., $\overline{J_x J_y}$=$J_x J_y$+$J_y J_x$.
}
\end{table}

Now we move on to the analysis of multipole properties
to clarify the ground-state properties from the viewpoint of multipole.
We measure multipole correlation functions
\begin{equation}
\chi_{\Gamma_{\gamma}}(q)=\sum_{j,k}
\langle X_{j\Gamma_{\gamma}} X_{k\Gamma_{\gamma}} \rangle
{\rm e}^{{\rm i}q(j-k)}/N,
\end{equation}
where $X_{i\Gamma_{\gamma}}$ is a multipole operator
with the symbol $X$ of multipole
for the irreducible representation $\Gamma_{\gamma}$
in the cubic symmetry at site $i$.
Here, we consider 15 types of multipoles
including three dipoles ($X$=$J$), five quadrupoles ($X$=$O$),
and seven octupoles ($X$=$T$),
as listed in Table I.
\cite{Shiina1997}
We evaluate the multipole correlation functions by DMRG calculations
with chains of 16 sites.

\begin{figure}[t]
\begin{center}
\includegraphics[width=0.95\linewidth]{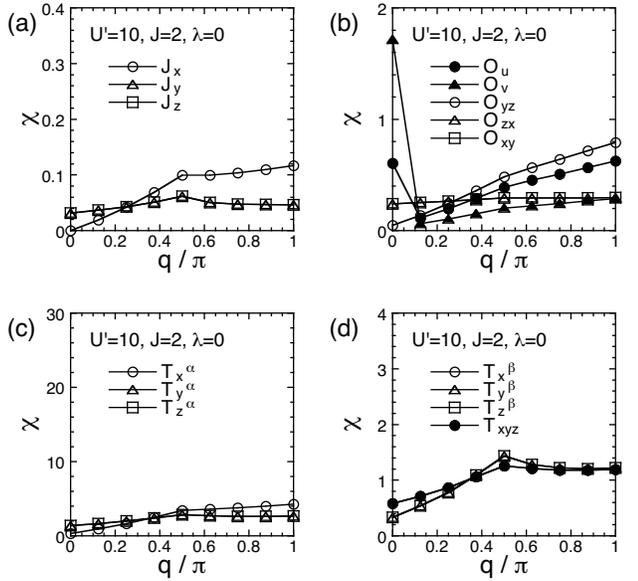}
\end{center}
\caption{%
DMRG results of multipole correlations
at $U'$=$10$, $J$=$2$, and $\lambda$=$0$:
(a) $\Gamma_{4u}$ dipoles;
(b) $\Gamma_{3g}$ and $\Gamma_{5g}$ quadrupoles;
(c) $\Gamma_{4u}$ octupoles;
and
(d) $\Gamma_{5u}$ and $\Gamma_{2u}$ octupoles.
}
\label{f2}
\end{figure}

Figure~2 shows DMRG results of the multipole correlation functions
at $U'$=$10$, $J$=$2$, and $\lambda$=$0$
for the paramagnetic phase.
Regarding dipoles, as shown in Fig.~2(a), 
the $J_{x}$ correlation has a peak at $q$=$\pi$,
which signals an antiferromagnetic state.
Here, we notice that each of the dipole correlations exhibits
a kink at $q$=$\pi/2$,
while the kink corresponds to a peak
for the $J_{y}$ and $J_{z}$ correlations.
This kink structure originates in the spin-orbital SU(4) symmetry
which realizes at a special point $J$=$\lambda$=$0$.
\cite{Yamashita1998,Lee2004,Xavier2006,Onishi2007a,Onishi2007b}
At the SU(4) symmetric point,
correlations of spin ${\bf S}_{i}$ and orbital pseudospin
${\bf T}_{i}$=$\frac{1}{2}\sum_{\tau,\tau',\sigma}
d_{i\tau\sigma}^{\dag}\mbox{\boldmath $\sigma$}_{\tau\tau'}d_{i\tau'\sigma}$,
where \mbox{\boldmath $\sigma$} are Pauli matrices,
coincide with each other
and have a peak at $q$=$\pi/2$.
With increasing $J$,
the spin correlation of $q$=$\pi/2$ grows
and the peak of the spin correlation remains at $q$=$\pi/2$.
On the other hand,
the pseudospin correlation of $q$=$\pi$ is enhanced,
and the peak position of the pseudospin correlation changes
to $q$=$\pi$.
Note that for the orbital angular momentum ${\bf L}_{i}$,
the correlation of the $q$=$\pi$ component is enhanced as well.
Thus, the orbital contribution to dipole
leads to the peak of the $J_{x}$ correlation at $q$=$\pi$
rather than $q$=$\pi/2$.

In Fig.~2(b),
we find that for the $O_{u}$ and $O_{v}$ correlations,
a sharp peak appears at $q$=$0$,
since $\langle O_{u} \rangle$ and $\langle O_{v} \rangle$
are turned out to be finite.
We also find a peak at $q$=$\pi$
for the $O_{u}$, $O_{v}$, and $O_{yz}$ correlations,
implying an antiferro-orbital state.
For all quadrupoles, there occurs a kink at $q$=$\pi/2$
in similar to the case of dipoles,
which is a trace of the SU(4) symmetry at $J$=$\lambda$=$0$.
As shown in Fig.~2(c) and 2(d),
we also observe a kink at $q$=$\pi/2$ for octupoles.

\begin{figure}[t]
\begin{center}
\includegraphics[width=0.95\linewidth]{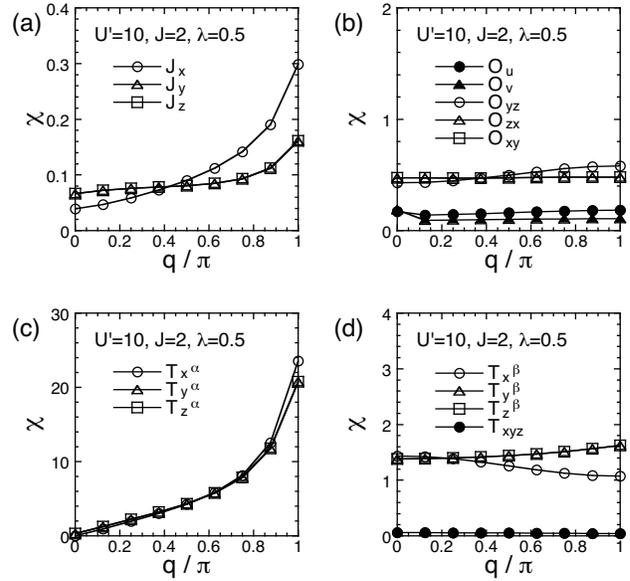}
\end{center}
\caption{%
DMRG results of multipole correlations
at $U'$=$10$, $J$=$2$, and $\lambda$=$0.5$:
(a) $\Gamma_{4u}$ dipoles;
(b) $\Gamma_{3g}$ and $\Gamma_{5g}$ quadrupoles;
(c) $\Gamma_{4u}$ octupoles;
and
(d) $\Gamma_{5u}$ and $\Gamma_{2u}$ octupoles.
}
\label{f3}
\end{figure}

In Fig.~3, we present the multipole correlation functions
at $U'$=$10$, $J$=$2$, and $\lambda$=$0.5$
for the ferromagnetic phase.
At a glance, we find that the kink structure at $q$=$\pi/2$
disappears for all multipoles.
Concerning dipoles, as shown in Fig.~3(a),
a peak appears at $q$=$\pi$.
Namely, antiferro-dipole correlations become dominant
even when the spin state is ferromagnetic
due to the orbital contribution.
In fact,
the spin ${\bf S}_{i}$ correlation has a peak at $q$=$0$,
while the orbital ${\bf L}_{i}$ correlation exhibits a peak at $q$=$\pi$
(not shown).
On the other hand,
the quadrupole correlations are found to be almost flat,
and we cannot see any fine structures signaling quadrupole ordering,
as shown in Fig.~3(b).
As for octupoles,
we observe a significant enhancement of
the $\Gamma_{4u}$ octupole correlations of the $q$=$\pi$ component
[see Figs.~2(c) and 3(c)].
The growth of the antiferro-octupole correlations reflects
the stabilization of
the antiferro-orbital state with complex orbitals.


In summary, we have studied the ground-state properites
of the $t_{\rm 2g}$-orbital Hubbard model with the spin-orbit coupling
from the viewpoint of multipole, by numerical techniques.
The strong spin-orbit coupling induces a transition
from the antiferromagnetic state to the ferromagnetic state.
We have found that
antiferro-dipole correlations develop
even when the spin state is ferromagnetic.
Moreover, the complex orbital state appears,
since the spin-orbit coupling yields the linear combinations
of spin and orbital states with complex number coefficients.
Accordingly, we observe an enhancement of
the $\Gamma_{4u}$ octupole correlations.
It is an interesting issue to explore possible multipole ordering
in $5d$-electron Ir compounds with strong spin-orbit coupling.


The author thanks G. Khaliullin, S. Maekawa, and M. Mori
for useful discussions.
This work was supported by Grant-in-Aid for Scientific Research
of Ministry of Education, Culture, Sports, Science, and Technoloty of Japan.


\end{document}